\documentclass[12pt]{article}
\usepackage{amsfonts}
\usepackage[mathscr]{eucal}
\usepackage[dvips]{color}

\newcommand{\be}{\begin{equation}}\newcommand{\ee}{\end{equation}}
\newcommand{\bea}{\begin{eqnarray}}
\newcommand{\eea}{\end{eqnarray}}

\newcommand{\p}[1]{(\ref{#1})}

\begin{document}

\begin{titlepage}

\vspace*{2cm}

\renewcommand{\thefootnote}{\star}
\begin{center}

{\LARGE\bf Superconformal Calogero models }

\vspace{0.5cm}

{\LARGE\bf  as a gauged matrix mechanics${}^\star$}

\vspace{2cm}

{\large\bf Sergey~Fedoruk}
\vspace{0.3cm}

{\it Bogoliubov  Laboratory of Theoretical Physics, JINR,}\\
{\it 141980 Dubna, Moscow region, Russia} \\
\vspace{0.1cm}

{\tt fedoruk@theor.jinr.ru}\\
\vspace{1cm}

\vspace{0.3cm}

\end{center}
\vspace{1cm} \vskip 0.6truecm  \nopagebreak

\begin{abstract}
\noindent
We present basics of the gauged superfield approach to
constructing ${\cal N}$-superconformal multi-particle Calogero-type systems
developed in arXiv:0812.4276, arXiv:0905.4951 and arXiv:0912.3508.
This approach is illustrated by the multi-particle systems
possessing ${\rm SU}(1,1|1)$ and $D(2,1;\alpha)$ supersymmetries,
as well as by the model of new ${\cal N}{=}4$ superconformal quantum mechanics.
\end{abstract}

\vspace{6cm}

\noindent ------------------------------------

\noindent ${}^\star$ \footnotesize{ Talk at the Conference ``Selected
Topics in Mathematical and Particle Physics'',  In Honor of 70-th Birthday of Jiri
Niederle, 5 - 7 May 2009, Prague and at the XVIII International Colloquium
``Integrable Systems and Quantum Symmetries'', 18 - 20 June 2009, Prague, Czech Republic.
}

\newpage

\end{titlepage}

\setcounter{footnote}{0}

\setcounter{equation}0
\section{Introduction}

The celebrated Calogero model~\cite{C} is a prime example
of an integrable and exactly solvable multi--particle system. It describes the system of
$n$ identical particles interacting through an inverse-square pair potential
$\sum\limits_{a\neq b} g/(x_a - x_b)^2$, $a,b=1,...,n$.
Calogero model and its generalizations provide deep connections of various branches of theoretical
physics and have a wide range of physical and mathematical applications
(for a review, see~\cite{OP,C2}).

An important property of the Calogero model is $d{=}1$ conformal symmetry ${\rm SO}(1,2)$.
Being multi--particle conformal mechanics, this model, in the two--particle case,
yields the standard conformal mechanics~\cite{AFF}. Conformal properties of
the Calogero model and supersymmetric generalizations of the latter
give possibilities to apply them in black hole physics,
since the near--horizon limits of the extreme black hole solutions
in $M$-theory correspond to $AdS_2$ geometry,
having the same ${\rm SO}(1,2)$ isometry group.
The analysis of physical fermionic degrees of freedom in the black hole solutions of
four- and five-dimensional supergravities shows that
related $d{=}1$ superconformal systems must possess ${\cal N}{=}\,4$ supersymmetry~\cite{BH,GT,MS}.

Superconformal Calogero models with ${\cal N}{=}\,2$ supersymmetry were considered in~\cite{FM,Vas}
and with ${\cal N}{=}\,4$ supersymmetry in~\cite{W,BGK,BGL,GLP,BKS,KLP}.
Unfortunately, a consistent Lagrange formulations for {$n$}-particle Calogero
model with ${\cal N}{=}\,4$ superconformal symmetry for any {$n$} is still lacking.

Recently, we developed a universal approach to superconformal Calogero models for an
arbitrary number
of interacting particles, including the ${\cal N}{=}\,4$ models.
It is based on the superfield gauging
of some non-abelian isometries of the $d{=}1$ field theories~\cite{DI}.

Our gauge model involves three matrix superfields. One is a bosonic superfield
in the adjoint representation of ${\rm U}(n)$. It carries physical degrees of freedom
of superCalogero system. The second superfield is in the fundamental (spinor) representation of
${\rm U}(n)$, and it is an auxiliary one and is described by Chern--Simons mechanical action~\cite{ChS,Poly0}.
The third matrix superfield accommodates gauge ``topological'' supermultiplet~\cite{DI}.
${\cal N}$-extended superconformal symmetry plays a very important role in our model.
Elimination of the pure gauge and auxiliary fields gives rise to Calogero--like
interactions for the physical fields.

The talk is based on the papers~\cite{FIL1,FIL2,FIL3}.

\setcounter{equation}0
\section{Gauged formulation of Calogero model}

The renowned Calogero system~\cite{C} can be  described by the following
action~\cite{Poly0,Gorsky}:
\begin{equation}\label{ac-bose}
S_0 = \int dt  \,\Big[\, { {\rm Tr}\left(\nabla\! X \nabla\! X \right)}  + {
{\textstyle\frac{i}{2}} (\bar Z \nabla\! Z - \nabla\! \bar Z Z)} + { c\,{\rm Tr} A
}\,\Big],
\end{equation}
where
$$
\nabla\! X = \dot X +i [A,X], \qquad \nabla\! Z = \dot Z + iAZ\, \qquad \nabla\! \bar Z =
\dot {\bar Z} - i\bar ZA\;.
$$
The action~(\ref{ac-bose}) is the action of ${\rm U}(n)$, $d{=}1$ gauge
theory. The hermitian $n{\times}n$-matrix field  ${ X_a^b(t)}$, $(\overline{X_a^b})
=X_b^a$, $a,b=1,\ldots ,n$ and complex commuting ${\rm U}(n)$-spinor field $Z_a(t)$, $\bar
Z^a = (\overline{Z_a})$ present the matter, scalar and spinor fields, respectively.
The $n^2$ ``gauge fields'' ${  A_a^b(t)}$, ${ (\overline{A_a^b}) =A_b^a}$ are
non--propagating ones in $d{=}1$ gauge theory.
The second term in the action~(\ref{ac-bose}) is the Wess--Zumino (WZ) term, whereas the third term is
the standard Fayet--Iliopoulos (FI) one.

The action~(\ref{ac-bose}) is invariant under the $d{=}1$ conformal ${\rm SO}(1,2)$
transformations:
\begin{equation}\label{ga-sl}
\delta t = \alpha, \qquad \delta X_a^b = {\textstyle\frac{1}{2}}\, \dot{\alpha}
X_a^b,\qquad \delta Z_a = 0,\qquad \delta A_a^b = -\dot{\alpha}A_a^b ,
\end{equation}
where constrained parameter $\partial_t^3 {\alpha} = 0$ contains three independent infinitesimal
constant parameters of ${\rm SO}(1,2)$.

The action~(\ref{ac-bose}) is also invariant with respects to
the local ${\rm U}(n)$ invariance
\begin{equation}\label{ga-u}
X \rightarrow  g X g^\dagger , \qquad  Z \rightarrow  g Z , \qquad A \rightarrow  g A
g^\dagger +i \dot g g^\dagger,
\end{equation}
where $g(\tau )\in {\rm U}(n)$.

Let us demonstrate, in Hamiltonian formalism, that the gauge model~(\ref{ac-bose})
is equivalent to the standard Calogero system.

The definitions of the momenta, corresponding to the action~(\ref{ac-bose}),
\begin{equation}\label{P-Cal}
P_{\!\scriptscriptstyle{X}} = 2 \nabla\! X\,,\qquad P_{\!\scriptscriptstyle{Z}} =
{\textstyle\frac{i}{2}} \, \bar Z \,,\qquad \bar P_{\!\scriptscriptstyle{Z}} =
-{\textstyle\frac{i}{2}}\, Z\,, \qquad P_{\!\scriptscriptstyle{A}} = 0
\end{equation}
imply the primary constraints
\begin{equation}\label{con-Z}
{\rm a)} \quad G \equiv P_{\!\scriptscriptstyle{Z}} - {\textstyle\frac{i}{2}} \, \bar Z
\approx 0\,,\quad \bar G \equiv \bar P_{\!\scriptscriptstyle{Z}} +
{\textstyle\frac{i}{2}}\, Z \approx 0\,;\qquad\qquad {\rm b)} \quad
P_{\!\scriptscriptstyle{A}} \approx 0
\end{equation}
and give us the following expression for the canonical Hamiltonian
\begin{equation}\label{H-Cal}
H = {\textstyle\frac{1}{4}}\, {\rm Tr}\left(
P_{\!\scriptscriptstyle{X}}P_{\!\scriptscriptstyle{X}} \right) - {\rm Tr}\left( A\, T
\right),
\end{equation}
where matrix quantity $T$ is defined as
\begin{equation}\label{con-T}
T \equiv \, i [X, P_{\!\scriptscriptstyle{X}} ] -  Z \!\cdot\! \bar Z + c I_n \,.
\end{equation}

The preservation of the constraints~(\ref{con-Z}b) in time leads to the secondary constraints
\begin{equation}\label{con-T1}
T \approx 0 \,.
\end{equation}
The gauge fields $A$ play the role of the Lagrange multipliers for these constraints.

Using canonical Poisson brackets $ [X_a^b, P_{\!\scriptscriptstyle{X}}{}_{c}^d ]_{{}_P}
{=}\delta_a^d \delta_c^b $, $[Z_a, P_{\!\scriptscriptstyle{Z}}^b ]_{{}_P} {=}\delta_a^b $,
$[{\bar Z}^a, \bar P_{{\!\scriptscriptstyle{Z}}\,b} ]_{{}_P} {=}\delta^a_b$, we obtain the Poisson
brackets of the constraints~(\ref{con-Z}a)
\begin{equation}\label{PB-G}
[G^a, \bar G_b ]_{{}_P} =-i\delta^a_b\,.
\end{equation}
Dirac brackets for these second class constraints~(\ref{con-Z}a) eliminates
spinor momenta $P_{\!\scriptscriptstyle{Z}}$, $\bar P_{\!\scriptscriptstyle{Z}}$ from the phase
space. The Dirac brackets for the residual variables take the form
\begin{equation}\label{CDB-Cal}
[X_a^b, P_{\!\scriptscriptstyle{X}}{}_{c}^d ]_{{}_D} =\delta_a^d \delta_c^b \,, \qquad
[Z_a, {\bar Z}^b ]_{{}_D} =-i\,\delta_a^b\,.
\end{equation}

The residual constraints~(\ref{con-T1})
$ T = T^+$ form $u(n)$ algebra with respect to the Dirac brackets
\begin{equation}\label{DB-T}
[T_a^b, T_c^d ]_{{}_D} =i(\delta_a^d T_c^b - \delta_c^b T_a^d )
\end{equation}
and generate gauge transformations~(\ref{ga-u}). Let us fix the gauges for these
transformations.

In the notations
$$
x_a \equiv X_a^a\,, \quad p_a \equiv P_{\!\scriptscriptstyle{X}}{}_{a}^a    \quad (\mbox{no
summation over } a)\,; \qquad x_a^b \equiv X_a^b\,, \quad p_a^b \equiv
P_{\!\scriptscriptstyle{X}}{}_{a}^b
  \quad \mbox{for } a\neq b
$$
the constraints~(\ref{con-T}) take the form
\begin{equation}\label{T-ab}
T_a^b =i(x_a -x_b) p_a^b - i(p_a -p_b) x_a^b + i\sum_{c}(x_a^c p_c^b - p_a^c x_c^b) - Z_a
{\bar Z}^b \approx 0  \qquad \mbox{for } a\neq b \,,
\end{equation}
\begin{equation}\label{T-a}
T_a^a =i\sum_{c}(x_a^c p_c^a - p_a^c x_c^a) - Z_a {\bar Z}^a + c \approx 0 \qquad (\mbox{no
summation over } a) \,.
\end{equation}
The non-diagonal constraints~(\ref{T-ab}) generate the transformations
$$
\delta x_a^b = [x_a^b, \epsilon_b^a T_b^a ]_{{}_D} \sim i(x_a -x_b)\epsilon_b^a\,.
$$
Therefore, in case of Calogero--like condition $x_a {\neq} x_b$, we can impose the gauge
\begin{equation}\label{gau-ab}
x_a^b \approx 0\,.
\end{equation}
Then we introduce Dirac brackets for the constraints~(\ref{T-ab}), (\ref{gau-ab}) and eliminate
$x_a^b$, $p_a^b$. In particular, the resolved expression for $p_a^b$
is
\begin{equation}\label{res-ab}
p_a^b = -\frac{i}{(x_a -x_b)}\, Z_a {\bar Z}^b\,.
\end{equation}
The Dirac brackets of residual variables coincide with Poisson ones due to the resolved form of
gauge fixing condition~(\ref{gau-ab}).

After gauge-fixing~(\ref{gau-ab}), the constraints~(\ref{T-a}) become
\begin{equation}\label{T-a-fix}
Z_a {\bar Z}^a - c \approx 0 \qquad (\mbox{no summation over } a)
\end{equation}
and generate local phase transformations of $Z_a$.
For these gauge transformations we impose the gauge
\begin{equation}\label{Z-fix}
Z_a - {\bar Z}^a \approx 0\,.
\end{equation}
The conditions~(\ref{T-a-fix}) and~(\ref{Z-fix}) eliminate $Z_a$ and ${\bar Z}^a$
completely.

Finally, using the expressions~(\ref{res-ab}) and the conditions~(\ref{gau-ab}), (\ref{T-a-fix}) we obtain
the following expression for the Hamiltonian~(\ref{H-Cal})
\begin{equation}\label{H-Cal-fix}
H_0 = {\textstyle\frac{1}{4}}\, {\rm Tr}\left(
P_{\!\scriptscriptstyle{X}}P_{\!\scriptscriptstyle{X}} \right) = \frac{1}{4}\left(\sum_{a}
(p_a )^2 + \sum_{a\neq b} \frac{c^2}{(x_a - x_b)^2}\right),
\end{equation}
which corresponds to the standard Calogero action~\cite{C}
\begin{equation}\label{A-Cal}
S_0 = \int dt  \,\Big[\, \sum_{a} \dot x_a \dot x_a - \sum_{a\neq b} \frac{c^2}{4(x_a -
x_b)^2}\,\Big].
\end{equation}

\setcounter{equation}0
\section{${\cal N}{=}\,2$ superconformal Calogero model}

${\cal N}{=}\,2$ supersymmetric generalization of the system~(\ref{ac-bose}) is described by
\begin{itemize}
\item
the even hermitian $(n\times n)$--matrix superfield \,\,$ \mathcal{X}_a^b(t,
\theta,\bar\theta)$, $(\mathcal{X})^+ =\mathcal{X}$, $a,b=1,\ldots ,n$ \qquad [the
supermultiplets ${\bf (1, 2, 1)}$];
\item
commuting chiral ${\rm U}(n)$--spinor superfield $ \mathcal{Z}_a
(t_{\!\scriptscriptstyle{L}}, \theta)$, $\bar \mathcal{Z}^a (t_{\!\scriptscriptstyle{R}},
\bar\theta) = (\mathcal{Z}_a)^+$, $t_{\!\scriptscriptstyle{L,R}}=t\pm i\theta\bar\theta$
\qquad [the supermultiplets ${\bf (2,2,0)}$];
\item
commuting $n^2$ complex ``bridge'' superfields $ b_a^c(t, \theta,\bar\theta)$.
\end{itemize}
${\cal N}{=}\,2$ superconformally invariant action of these superfields has the form
\begin{equation}\label{N2-Cal-v}
S_{2} = \int dt d^2\theta \,\Big[\, {\rm Tr} \left( \bar\mathcal{D} \mathcal{X} \,
\mathcal{D} \mathcal{X}\, \right) + {\textstyle\frac12}\,\bar \mathcal{Z}\, e^{2V}\!
\mathcal{Z} - c\,{\rm Tr} V \,\Big] .
\end{equation}
Here the covariant derivatives of the superfield $\mathscr{X}$ are
\begin{equation}\label{cov-der-v}
\mathcal{D} \mathcal{X} =  D \mathcal{X} +i [ \mathscr{A} , \mathcal{X}] \,, \qquad
\bar\mathcal{D} \mathcal{X} = \bar D \mathcal{X} +i [ \bar\mathscr{A} , \mathcal{X}]\,,
\end{equation}
$$
D = \partial_{\theta} +i\bar\theta\partial_{t}\,, \quad \bar D = -\partial_{\bar\theta}
-i\theta\partial_{t}\,, \quad \{D, \bar D \} = -2i \partial_{t}\,,
$$
where the potentials are constructed from the bridges as
\begin{equation}\label{pot}
\mathscr{A} =  -i \, e^{i\bar b} (D e^{-i\bar b}) \,,  \qquad \bar\mathscr{A} = -i \, e^{ib}
(\bar D e^{-ib}) \qquad\qquad (\bar b \equiv b^+)\,.
\end{equation}
The gauge superfield prepotential $V_a^b(t,\theta,\bar\theta)$, $(V)^\dagger = V$, is
constructed from the bridges as
\begin{equation}\label{prepot}
e^{2V} = e^{-i\bar b} \, e^{ib} \,.
\end{equation}

The superconformal boosts  of the ${\cal N}{=}\,2$ superconformal group ${\rm
SU}(1,1|1)\simeq {\rm OSp}(2|2)$ have the following
realization:
\begin{equation}  \label{2N-sc-c}
\delta t= -i(\eta\bar\theta+\bar\eta\theta)t\,,\qquad \delta \theta =
\eta(t+i\theta\bar\theta)\,,\qquad \delta \bar\theta = \eta(t-i\theta\bar\theta)\,,
\end{equation}
\begin{equation}  \label{2N-sc-X}
\delta \mathcal{X}= -i(\eta\bar\theta+\bar\eta\theta)\,\mathcal{X}\,, \qquad \delta
\mathcal{Z} = 0\,,\qquad \delta b = 0\,,\qquad \delta V = 0\,.
\end{equation}
Its closure with  ${\cal N}{=}\,2$ supertranslations yields the full ${\cal N}{=}\,2$
superconformal invariance of the action~(\ref{N2-Cal-v}).

The action~(\ref{N2-Cal-v}) is invariant also with respect to the two types of
the local ${\rm U}(n)$ transformations:\\
$\bullet$ $\tau$--transformations with the hermitian $(n\times n)$--matrix parameter
$\tau (t, \theta,\bar\theta) \in u(n)$, $(\tau)^+ =\tau$;\\
$\bullet$ $\lambda$--transformations with complex chiral gauge parameters $\lambda
(t_{\!\scriptscriptstyle{L}}, \theta) \in u(n)$, $\bar\lambda (t_{\!\scriptscriptstyle{R}},
\theta) = (\lambda)^+ $.\\
These ${\rm U}(n)$  transformations act on the superfields in the action~(\ref{N2-Cal-v}) as
\begin{equation}\label{tran-b}
e^{ib^\prime} = e^{i\tau} \, e^{ib} e^{-i\lambda}\,, \qquad e^{2V^\prime} =
e^{i\bar\lambda} \, e^{2V} e^{-i\lambda}\,,
\end{equation}
\begin{equation}\label{tran-X}
\mathcal{X}^{\,\prime} =  e^{i\tau}\, \mathcal{X}\, e^{-i\tau} \,, \qquad
\mathcal{Z}^{\prime} =  e^{i\lambda} \mathcal{Z} \,, \qquad \bar \mathcal{Z}^{\prime} =
\bar \mathcal{Z}\, e^{-i\bar\lambda}\,.
\end{equation}

In terms of $\tau$--invariant superfields $V$, $\mathcal{Z}$
and new hermitian $(n\times n)$--matrix superfield
\begin{equation}\label{X-tX}
\mathscr{X} = e^{-ib} \,\mathcal{X}\, e^{i\bar b}\,,\qquad
\mathscr{X}^{\,\prime} =  e^{i\lambda}\, \mathscr{X}\, e^{-i\bar\lambda} \,,
\end{equation}
the action~(\ref{N2-Cal-v}) takes the form
\begin{equation}\label{N2-Cal}
S_{2} = \int dt d^2\theta \,\Big[\, {\rm Tr} \left( \bar\mathscr{D} \mathscr{X} \, e^{2V}
\mathscr{D} \mathscr{X}\, e^{2V} \right) + {\textstyle\frac12}\,\bar \mathcal{Z}\, e^{2V}\!
\mathcal{Z} - c\,{\rm Tr} V \,\Big]
\end{equation}
where the covariant derivatives of the superfield $\mathscr{X}$ are
\begin{equation}\label{cov-der-s2}
\mathscr{D} \mathscr{X} =  D \mathscr{X} + e^{-2V} (D e^{2V}) \, \mathscr{X} \,, \qquad
\bar\mathscr{D} \mathscr{X} = \bar D \mathscr{X} - \mathscr{X} \, e^{2V} (\bar D
e^{-2V})\,.
\end{equation}

For gauge $\lambda$--transformations we impose WZ gauge
$$
V (t, \theta,\bar\theta) = -\theta\bar\theta A (t)\,.
$$
Then, the action~(\ref{N2-Cal}) takes the form
\begin{equation}\label{N2-Cal-c}
S_{2} = { S_{0}}+ {S^{\Psi}_{2}}, \qquad { S^{\Psi}_{2}= -i\, {\rm Tr} {\displaystyle\int}
dt \,( \bar\Psi \nabla \Psi - \nabla \bar\Psi \Psi )}
\end{equation}
where $\Psi = D \mathscr{X}|$ and
$$
\nabla \Psi = \dot \Psi +i [A,\Psi]\,, \qquad \nabla \bar\Psi = \dot {\bar\Psi} +i
[A,\bar\Psi]\,.
$$
The bosonic core in~(\ref{N2-Cal-c}) exactly coincides with the Calogero action~(\ref{A-Cal}).

Exactly as in pure bosonic case, residual local U(n) invariance
of the action~(\ref{N2-Cal-c}) eliminates the nondiagonal fields $X_a^b$,
$a{\neq}b$, and all spinor fields $Z_a$. Thus, the physical fields
in our ${\cal N}{=}\,2$ supersymmetric generalization of the Calogero system are $n$ bosons
$x_a=X_a^a$ and $2n^2$ fermions $\Psi_a^b$. These fields present on--shell content
of $n$ multiplets {\bf (1,2,1)} and $n^2{-}n $
multiplets {\bf (0,2,2)} which are obtained from $n^2$ multiplets {\bf (1,2,1)}
by gauging procedure~\cite{DI}. We can present it by the
plot:
\begin{center}
$\underbrace{\mathscr{X}_a^a=({X}_a^a, \Psi_a^a, C_a^a)}_{{{\bf (1,2,1)}\,multiplets}}$\qquad\quad
\qquad $\underbrace{\mathscr{X}_a^b=({{X}_a^b}, \Psi_a^b, C_a^b),
{\scriptstyle a\neq b}}_{{\bf (1,2,1)}\,multiplets}$\\
$\,$\\
$\,$\\
{$\Downarrow$ \qquad{gauging}\qquad $\Downarrow$}\quad$\,$\\
$\,$\\
$\,$\\
$\underbrace{\mathscr{X}_a^a=({X}_a^a, \Psi_a^a, C_a^a)}_{{\bf (1,2,1)}\,multiplets}$\,\,\quad
{interact}\quad $\underbrace{\Omega_a^b=(\Psi_a^b, {B_a^b}, C_a^b), {\scriptstyle a\neq b}}_{{\bf
(0,2,2)}\,multiplets}$
\end{center}

\noindent where the bosonic fields $C_a^a$, $C_a^b$ and $B_a^b$
are auxiliary components of the supermultiplets. Thus, we obtain some new ${\cal
N}{=}\,2$ extensions of the $n$-particle Calogero models with {$n$} bosons and {$2n^2$}
fermions as compared to the standard ${\cal N}{=}\,2$ superCalogero with {$2n$} fermions
constructed by Freedman and Mende~\cite{FM}.

\setcounter{equation}0
\section{${\cal N}{=}\,4$ superconformal Calogero model}

The most natural formulation of ${\cal N}{=}\,4, d{=}\,1$ superfield theories is achieved in the
harmonic superspace~\cite{IL} parametrized by
$$
(t,\theta_i, \bar\theta^k, u_i^\pm)\sim
(t,\theta^\pm, \bar\theta^\pm, u_i^\pm)\,, \qquad
\theta^\pm=\theta^i u_i^\pm,\quad \bar\theta^\pm=\bar\theta^i u_i^\pm\,,\qquad i,k =1,2.
$$
Commuting ${\rm SU}(2)$-doublets $u_i^\pm$ are harmonic coordinates~\cite{GIOS}, subjected by the constraints
$u^{+i}u_i^-=1$. The ${\cal N}{=}\,4$ superconformally invariant
harmonic analytic subspace is parametrized by
$$
(\zeta,u)=(t_A,\theta^+, \bar\theta^+, u_i^\pm), \qquad t_A=t-i(\theta^+ \bar\theta^-
+\theta^-\bar\theta^+)\,.
$$
The integration measures in these superspaces are
$\mu_H =dudtd^4\theta$ and $\mu^{(-2)}_A=dud\zeta^{(-2)}$.

${\cal N}{=}\,4$ supergauge theory related to our task is described by:
\begin{itemize}
\item  hermitian matrix superfields ${\mathscr{X}(t,\theta^\pm, \bar\theta^\pm, u_i^\pm)
=(\mathscr{X}_a^b)}$ subjected to the
constraints
\begin{equation}\label{N4-Xcon}
\mathscr{D}^{++} \,\mathscr{X}=0, \qquad \mathscr{D}^{+}\mathscr{D}^{-} \,\mathscr{X}=0,
\qquad  (\mathscr{D}^{+}\bar{\mathscr{D}}^{-} +\bar{\mathscr{D}}^{+}\mathscr{D}^{-})\,
\mathscr{X}=0
\end{equation}
[the multiplets {\bf (1,4,3)}];
\item  analytic superfields $\mathcal{Z}^+(\zeta,u)=(\mathcal{Z}^+_a)$ subjected to the
constraint
\begin{equation}\label{N4-Zcon}
\mathscr{D}^{++} \mathcal{Z}^+=0
\end{equation}
[the multiplets {\bf (4,4,0)}];
\item  the gauge matrix connection ${V^{++}(\zeta,u)}=(V^{++}{}_a^b)$.
\end{itemize}
In~(\ref{N4-Xcon}) and~(\ref{N4-Zcon}) covariant derivatives
are defined by
$$
\mathscr{D}^{++} \mathscr{X} = D^{++}
\mathscr{X} + i\,[V^{++} ,\mathscr{X}] , \qquad \mathscr{D}^{++} \mathcal{Z}^+ = D^{++}
\mathcal{Z}^+ + i\,V^{++} \mathcal{Z}^+.
$$
Also ${\mathscr{D}}^{+}=D^+$, $\bar{\mathscr{D}}^{+}=\bar D^+$ and the connections in
${\mathscr{D}}^{-}$, $\bar{\mathscr{D}}^{-}$ are expressed through derivatives of~$V^{++}$.

The ${\cal N}{=}\,4$ superconformal model is described by the action
\begin{equation}\label{4N-gau-matrix}
S_4^{\alpha\neq 0} = -{\textstyle\frac{1}{4(1+\alpha)}}\int \mu_H\,  {\rm Tr} \left(
\mathscr{X}^{\,-1/\alpha} \, \right) + {\textstyle\frac{1}{2}}\int \mu^{(-2)}_A\,
\mathcal{V}_0 \,\widetilde{\mathcal{Z}}{}^+ \mathcal{Z}^+ + {\textstyle\frac{i}{2}}\,c\int
\mu^{(-2)}_A \,{\rm Tr} \,V^{++} \,.
\end{equation}
The tilde in $\widetilde{\mathcal{Z}}{}^+$ denotes `hermitian' conjugation
preserving analyticity~\cite{GIOS,IL}.

The unconstrained superfield $\mathcal{V}_0(\zeta,u)$ is a real analytic superfield, which is defined by
the integral transform ($\mathscr{X}_0 \equiv {\rm Tr} \left( \mathscr{X} \right)$)
$$
\mathscr{X}_0(t,\theta_i,\bar\theta^i)=\int du\, \mathcal{V}_0 \left(t_A, \theta^+,
\bar\theta^+, u^\pm \right) \Big|_{\theta^\pm=\theta^i u^\pm_i,\,\,\,
\bar\theta^\pm=\bar\theta^i u^\pm_i}.
$$

The real number $\alpha{\neq} 0$ in~(\ref{4N-gau-matrix}) coincides with the parameter of
${\cal N}{=}\,4$ superconformal group $D(2,1;\alpha)$ which is symmetry group of the
action~(\ref{4N-gau-matrix}). Field transformations under superconformal boosts are (see the
coordinate transformations in~\cite{IL,DI})
\begin{equation}\label{4N-sc-f}
\delta \mathscr{X}= -\Lambda_0\,\mathscr{X}\,,\qquad \delta \mathcal{Z}^+ =
\Lambda\,\mathcal{Z}^+,\qquad \delta V^{++} = 0\,,
\end{equation}
where $\Lambda = 2i\alpha(\bar\eta^-\theta^+ - \eta^-\bar\theta^+)$, $\Lambda_0 = 2\Lambda-
D^{--} D^{++}\Lambda$.
It is important that just the superfield multiplier $\mathcal{V}_0$
in the action provides this invariance due to $\delta\mathcal{V}_0 = -2\Lambda
\mathcal{V}_0$ (note that $\delta \mu^{(-2)}_A= 0$).

The action~(\ref{4N-gau-matrix}) is invariant under the local ${\rm U}(n)$ transformations:
\begin{equation}\label{4N-un-f}
\mathscr{X}^{\,\prime} =  e^{i\lambda} \mathscr{X} e^{-i\lambda} , \qquad
\mathcal{Z}^+{}^{\prime} = e^{i\lambda} \mathcal{Z}^+, \qquad V^{++}{}^{\,\prime} =
e^{i\lambda}\, V^{++}\, e^{-i\lambda} - i\, e^{i\lambda} (D^{++} e^{-i\lambda}) ,
\end{equation}
where $ \lambda_a^b(\zeta, u^\pm) \in u(n) $ is the `hermitian' analytic matrix parameter,
$\widetilde{\lambda} =\lambda$. Using gauge freedom~(\ref{4N-un-f}) we choose the WZ gauge
\begin{equation}\label{4N-WZ}
V^{++} =-2i\,\theta^{+}
 \bar\theta^{+}A(t_A) .
\end{equation}
Considering the case $\alpha{=}-\frac12$ (when $D(2,1;\alpha)\simeq {\rm OSp}(4|2)$) in the
WZ gauge and eliminating auxiliary and gauge fields, we find that the
action~(\ref{4N-gau-matrix}) has the following bosonic limit
\begin{equation}\label{4N-ac-bo}
S_{4,b}^{\alpha{=}-1/2} = \int dt \left\{ \sum_{a} \dot x_a \dot x_a +
{\textstyle\frac{i}{2}}\sum_{a} (\bar Z_k^a \dot Z^k_a - \dot {\bar Z}{}_k^a Z^k_a)  +
\sum_{a\neq b} \frac{{\rm Tr}(S_a S_b)}{4(x_a - x_b)^2} - \frac{n\,{\rm Tr}(\hat S \hat
S)}{2(X_0)^2}\right\},
\end{equation}
where
$$
(S_a)_i{}^j \equiv \bar Z^a_i Z_a^j, \qquad (\hat S)_i{}^j \equiv \sum_a \left[ (S_a)_i{}^j
- {\textstyle\frac{1}{2}}\delta_i^j(S_a)_k{}^k\right]\,.
$$
The fields $x_a$ are ``diagonal'' fields in $X=\mathscr{X}|$. The fields $Z^{i}$ define
first components in $\mathcal{Z}^+$, $\mathcal{Z}^+| = Z^{i}u_i^+ $. They are subject to the constraints
\begin{equation}\label{4N-Z-con}
\bar Z_i^a Z^i_a =c \qquad \forall \, a \,.
\end{equation}
These constraints are generated by the equations of motion with respect to the diagonal components
of gauge field $A$.

Using Dirac brackets $[\bar Z^a_i, Z_b^j]_{{}_D}= i\delta^a_b\delta_i^j$, which are generated
by the kinetic WZ term for $Z$, we find that the quantities $S_a$ for each $a$ form $u(2)$ algebras
$$
[(S_a)_i{}^j, (S_b)_k{}^l]_{{}_D}= i\delta_{ab}\left\{\delta_i^l(S_a)_k{}^j-
\delta_k^j(S_a)_i{}^l \right\}.
$$
Thus modulo center-of-mass conformal potential (up to the last term in~(\ref{4N-ac-bo})),
the bosonic limit~(\ref{4N-ac-bo}) is none other than the integrable U(2)-spin
Calogero model in the formulation of~\cite{Poly1,C2}.
Except for the case $\alpha{=}-\frac12$,
the action~(\ref{4N-gau-matrix})  yields
non--trivial sigma--model type kinetic term for the field $X=\mathscr{X}|$.

For $\alpha{=}0$ it is necessary to modify the transformation law of $\mathscr{X}$ in the
following way~\cite{DI}
\begin{equation}  \label{sc-X-m}
\delta_{mod} \mathscr{X}=2i(\theta_k\bar\eta^k + \bar\theta^k\eta_k) \,.
\end{equation}
Then the $D(2,1;\alpha{=}0)$ superconformal action reads
\begin{equation}\label{4N-X-0}
S_{4}^{\alpha= 0}
 =  -{\textstyle\frac{1}{4}} \int \mu_H \,  {\rm Tr} \left( e^{\,\mathscr{X}} \right)+
{\textstyle\frac{1}{2}}\int \mu^{(-2)}_A\, \widetilde{\mathcal{Z}}{}^+ \mathcal{Z}^+ +
{\textstyle\frac{i}{2}}\,c\int \mu^{(-2)}_A \,{\rm Tr} \, V^{++} \, .
\end{equation}
The $D(2,1;\alpha{=}0)$ superconformal invariance is not compatible with the presence of
$\mathcal{V}$ in the WZ term of the action~(\ref{4N-X-0}), still implying the
transformation laws~(\ref{4N-sc-f}) for $\mathcal{Z}^+$ and for~$V^{++}\,$. This situation
is quite analogous to what happens in the ${\cal N}{=}2$ super Calogero
model considered in Sect. 3, where the center-of-mass supermultiplet ${\rm Tr}(\mathscr{X})$
decouples from the WZ and gauge supermultiplets. Note that the (matrix) $\mathscr{X}$
supermultiplet interacts with the (column) $\mathcal{Z}$ supermultiplet
in~(\ref{N2-Cal-v}) and~(\ref{4N-X-0}) via the gauge supermultiplet.

\setcounter{equation}0
\section{$D(2,1;\alpha)$ quantum mechanics}

The $n{=}1$ case of the ${\cal N}{=}4$ Calogero--like model~(\ref{4N-gau-matrix}) above
 (the center-of-mass coordinate case) amounts to a non-trivial model  of ${\cal N}{=}4$
superconformal mechanics.

Choosing WZ gauge~(\ref{4N-WZ}) and eliminating the auxiliary fields by their algebraic
equations of motion, we obtain that the action takes the following on-shell form
\begin{eqnarray}
S &=& S_b+ S_f\,, \label{4N-ph}\\
S_b &=&  \int dt \,\Big[\dot x\dot x  + {\textstyle\frac{i}{2}} \left(\bar z_k \dot z^k -
\dot{\bar z}_k z^k\right)-\frac{\alpha^2(\bar z_k
z^{k})^2}{4x^2} -A \left(\bar z_k z^{k} -c \right) \Big] \,,\label{bose}\\
S_f &=&  -i \!\int\! dt \left( \bar\psi_k \dot\psi^k -\dot{\bar\psi}_k \psi^k \right) +
2\alpha  \!\int\! dt \, \frac{\psi^{i}\bar\psi^{k} z_{(i} \bar z_{k)}}{x^2} +
\textstyle{\frac{2}{3}}\,(1+2\alpha) \!\displaystyle{\int}\! dt\,
\frac{\psi^{i}\bar\psi^{k} \psi_{(i}\bar\psi_{k)}}{x^2}  . \label{fermi}
\end{eqnarray}

The action~(\ref{4N-ph}) possesses $D(2,1;\alpha)$ superconformal invariance. Using the
N\"other procedure, we find the $D(2,1;\alpha)$ generators. Quantum counterpart of them are
\begin{equation}\label{Q-qu}
\mathbf{Q}^i =P \Psi^i+ 2i\alpha\frac{Z^{(i} \bar Z^{k)}\Psi_k}{X}+
i(1+2\alpha)\frac{\langle\Psi_{k} \Psi^{k}\bar\Psi^i\rangle}{X}\, ,
\end{equation}
\begin{equation}\label{Qb-qu}
\bar\mathbf{Q}_i=P \bar\Psi_i- 2i\alpha\frac{Z_{(i} \bar Z_{k)}\bar\Psi^k}{X} +
i(1+2\alpha)\frac{\langle\bar\Psi^{k} \bar\Psi_{k}\Psi_i\rangle}{X}\,,
\end{equation}
\begin{equation}\label{S-qu}
\mathbf{S}^i =-2\,X \Psi^i + t\,\mathbf{Q}^i,\qquad \bar\mathbf{S}_i=-2\,X
\bar\Psi_i+ t\,\bar\mathbf{Q}_i\,.
\end{equation}
\begin{equation}\label{H-qu}
\mathbf{H} ={\textstyle\frac{1}{4}}\,P^2  +\alpha^2\frac{(\bar Z_k Z^{k})^2+2\bar Z_k
Z^{k}}{4X^2} - 2\alpha    \frac{Z^{(i} \bar Z^{k)} \Psi_{(i}\bar\Psi_{k)}}{X^2}  -\,
(1+2\alpha) \frac{\langle\Psi_{i}\Psi^{i}\,\bar\Psi^{k} \bar\Psi_{k}\rangle}{2X^2}+
  \frac{(1+2\alpha)^2}{16X^2}\,,
\end{equation}
\begin{equation}\label{K-qu}
\mathbf{K} =X^2  - t\,{\textstyle\frac{1}{2}}\,\{X, P\} +
    t^2\, \mathbf{H}\,,
\qquad \mathbf{D} =-{\textstyle\frac{1}{4}}\,\{X, P\} +
    t\, \mathbf{H}\,,
\end{equation}
\begin{equation}\label{I-qu}
\mathbf{J}^{ik} = i\left[ Z^{(i} \bar Z^{k)}+ 2\Psi^{(i}\bar\Psi^{k)}\right]\,, \quad
\mathbf{I}^{1^\prime 1^\prime} = -i\Psi_k\Psi^k\,,\quad \mathbf{I}^{2^\prime 2^\prime} =
i\bar\Psi^k\bar\Psi_k\,,\quad \mathbf{I}^{1^\prime 2^\prime} =-{\textstyle\frac{i}{2}}\,
[\Psi_k,\bar\Psi^k]\,.
\end{equation}
The symbol $\langle...\rangle$ denotes Weyl ordering.

It can be directly checked that the generators~(\ref{Q-qu})--(\ref{I-qu}) form the
$D(2,1;\alpha)$ superalgebra
\begin{equation}
\{\mathbf{Q}^{ai^\prime i}, \mathbf{Q}^{bk^\prime k}\}=
-2\left(\epsilon^{ik}\epsilon^{i^\prime k^\prime} \mathbf{T}^{ab}+\alpha
\epsilon^{ab}\epsilon^{i^\prime k^\prime} \mathbf{J}^{ik}-(1+\alpha)
\epsilon^{ab}\epsilon^{ik} \mathbf{I}^{i^\prime k^\prime}\right)\,,\label{qB-Q-g}
\end{equation}
\begin{equation}
[\mathbf{T}^{ab}, \mathbf{T}^{cd}]= -i\left(\epsilon^{ac}\mathbf{T}^{bd}
+\epsilon^{bd}\mathbf{T}^{ac}\right)\,,
\end{equation}
\begin{equation}
[\mathbf{J}^{ij}, \mathbf{J}^{kl}]= -i\left(\epsilon^{ik}\mathbf{J}^{jl}
+\epsilon^{jl}\mathbf{J}^{ik}\right)\,,\qquad [\mathbf{I}^{i^\prime j^\prime},
\mathbf{I}^{k^\prime l^\prime}]= -i\big(\epsilon^{ik}\mathbf{I}^{j^\prime l^\prime}
+\epsilon^{j^\prime l^\prime}\mathbf{I}^{i^\prime k^\prime}\big)\,,\label{qB-J}
\end{equation}
\begin{equation}
[\mathbf{T}^{ab}, \mathbf{Q}^{ci^\prime i}]=i\epsilon^{c(a}\mathbf{Q}^{b)i^\prime i}
,\qquad [\mathbf{J}^{ij}, \mathbf{Q}^{ai^\prime k}]=i\epsilon^{k(i}\mathbf{Q}^{ai^\prime
j)},\qquad [\mathbf{J}^{i^\prime j^\prime}, \mathbf{Q}^{ak^\prime i}]=i\epsilon^{k^\prime
(i^\prime}\mathbf{Q}^{aj^\prime ) i} \label{qB-JQ}
\end{equation}
due to the quantum brackets
\begin{equation}\label{cB}
[X, P] = i\,, \qquad [Z^i, \bar Z_j] = \delta^i_j \,, \qquad \{\Psi^i, \bar\Psi_j\}=
-{\textstyle\frac{1}{2}}\,\delta^i_j \,.
\end{equation}
In~(\ref{qB-Q-g})-(\ref{qB-JQ}) we use the notation \,\, $\mathbf{Q}^{21^\prime
i}=-\mathbf{Q}^{i}$, $\mathbf{Q}^{22^\prime i}=-\bar\mathbf{Q}^{i}$, \quad
$\mathbf{Q}^{11^\prime i}=\mathbf{S}^{i}$, $\mathbf{Q}^{12^\prime i}=\bar\mathbf{S}^{i}$,
$\mathbf{T}^{22}=\mathbf{H}$, $\mathbf{T}^{11}=\mathbf{K}$, $\mathbf{T}^{12}=-\mathbf{D}$.

To find the quantum spectrum, we make use of the
realization
\begin{equation}\label{bo-re-Z}
\bar Z_i=v^+_i, \qquad Z^i=  \partial/\partial v^+_i
\end{equation}
for the bosonic operators where  $v^+_i$
 is a commuting complex ${\rm SU}(2)$ spinor, as well as the following realization of
the odd operators
\begin{equation}\label{q-re-Psi}
\Psi^i=\psi^i, \qquad \bar\Psi_i= -{\textstyle\frac{1}{2}}\,
\partial/\partial\psi^i\,,
\end{equation}
where $\psi^i$ are complex Grassmann variables.

The full wave function $\Phi=A_{1}+ \psi^i B_i +\psi^i\psi_i A_{2}$ is subjected to the
constraints
\begin{equation}\label{q-con}
\bar Z_i Z^i \,\Phi=v^+_i\frac{\partial}{\partial v^+_i}\,\Phi=c\,\Phi.
\end{equation}
Requiring the wave function $\Phi(v^+)$ to be single-valued gives rise to the condition
that positive constant $c$ is integer, $c\in \mathbb{Z}$. Then \p{q-con} implies that the
wave function $\Phi(v^+)$ is a homogeneous polynomial in $v^+_i$ of the degree $c$:
\begin{equation}\label{w-f-d}
\Phi=A^{(c)}_{1}+ \psi^i B^{(c)}_i +\psi^i\psi_i A^{(c)}_{2} \,,
\end{equation}
\begin{equation}\label{A-irred}
A^{(c)}_{i^\prime} = A_{i^\prime,}{}_{k_1\ldots k_{c}}v^{+k_1}\ldots v^{+k_{c}} \,,
\end{equation}
\begin{equation}\label{B-irred}
B^{(c)}_i = B^{\prime(c)}_i +B^{\prime\prime(c)}_i=v^+_i B^\prime_{k_1\ldots
k_{c-1}}v^{+k_1}\ldots v^{+k_{c-1}} + B^{\prime\prime}_{(ik_1\ldots k_{c})}v^{+k_1}\ldots
v^{+k_{c}}\,.
\end{equation}

On the physical states~(\ref{q-con}), (\ref{w-f-d})  Casimir operator
takes the value
\begin{equation}\label{qu-Cas-ev}
\mathbf{C}_2= \mathbf{T}^2 +\alpha\, \mathbf{J}^2- (1+\alpha)\,\mathbf{I}^2 +
{\textstyle\frac{i}{4}}\, \mathbf{Q}^{ai^\prime i}\mathbf{Q}_{ai^\prime i}=
\alpha(1+\alpha)(c+1)^2/4 \, .
\end{equation}

On the same states, the Casimir operators of the bosonic
subgroups ${\rm SU}(1,1)$, ${\rm SU(2)}_R$ and ${\rm SU(2)}_L$,
$$
\mathbf{T}^2=r_0(r_0-1)\,, \qquad \mathbf{J}^2=j(j+1)\,, \qquad \mathbf{I}^2=i(i+1)\,,
$$
take the values listed in the Table
\begin{table}[h]
\begin{center}
\renewcommand{\arraystretch}{2}
\begin{tabular}{|c|c|c|c|}
\hline & $r_0$ & $j$ & $i$ \\ \hline
     $A^{(c)}_{k^\prime}(x,v^+)$ & $\frac{|\alpha|(c+1)+1}{2}$ & $\frac{c}{2}$ &
$\frac{1}{2}$ \\
\hline
     $B^{\prime(c)}_{k}(x,v^+)$ & $\frac{|\alpha|(c+1)+1}{2} - \frac{1}{2}\,\rm{sign}(\alpha)$ & $\frac{c}{2}
-\frac{1}{2}$ & 0 \\ \hline
     $B^{\prime\prime(c)}_{k}(x,v^+)$ & $\frac{|\alpha|(c+1)+1}{2} + \frac{1}{2}\,\rm{sign}(\alpha)$ &
     $\frac{c}{2}+ \frac{1}{2}$& 0 \\ \hline
\end{tabular}\\
\end{center}
\end{table}
\\
The fields $B^{\prime}_i$ and $B^{\prime\prime}_{i}$ form doublets of SU(2)$_R$ generated
by $\mathbf{J}^{ik}\,$, whereas the component fields $A_{i^\prime}=(A_{1},A_{2})$ form a
doublet of ${\rm SU(2)}_L$ generated by $\mathbf{I}^{i^\prime k^\prime}$.

Each of $A_{i^\prime}$, $B^{\prime}_i$, $B^{\prime\prime}_{i}$  carries a representation of
the SU(1,1) group. Basis functions of these representations are eigenvectors of the
generator
 $
{\mathbf{R}}={\textstyle\frac{1}{2}}\,\left(a^{-1}{\mathbf{K}}+ a{\mathbf{H}}\right), $
where $a$ is a constant of the length dimension. These eigenvalues are $r=r_0 +n$, $n\in
\mathbb{N}$.

\setcounter{equation}0
\section{Outlook}

In~\cite{FIL1,FIL2,FIL3}, we proposed a new gauge approach to the
construction of superconformal Calogero-type systems. The characteristic features of this
approach are the presence of auxiliary supermultiplets with WZ type actions, the built-in
superconformal invariance and the emergence of the Calogero coupling constant as a strength
of the FI term of the U(1) gauge (super)field.

We see continuation of the researches presented in the solution of some problems, such as
\begin{itemize}
\item An analysis of possible integrability properties of new superCalogero models
with finding-out a role of the contribution of the center of mass in the case of
$D(2,1;\alpha)$, $\alpha{\neq}0$, invariant systems.
\item Construction of quantum ${\cal N}{=}4$ superconformal Calogero systems
by canonical quantization of systems~(\ref{4N-gau-matrix}) and~(\ref{4N-X-0}).
\item Obtaining the systems, constructed from mirror supermultiplets
and possessing $D(2,1;\alpha)$ symmetry, after use gauging procedures
in bi-harmonic superspace~\cite{IN}.
\item Obtaining other superextensions of the Calogero model distinct from the {$A_{n-1}$} type
(related to the root system of ${\rm SU}(n)$ group),
by applying the gauging procedure to other gauge groups.
\end{itemize}

\bigskip

\noindent {\bf\large Acknowledgements}\\

\noindent
I thank the Organizers of Jiri Niederle's Fest and the XVIII International Colloquium
for the kind hospitality in Prague. I would also like to thank my co-authors E.~Ivanov and O.~Lechtenfeld
for a fruitful collaboration.
I acknowledge a support from the RFBR grants 08-02-90490, 09-02-01209 and
09-01-93107 and grants
of the Heisenberg-Landau and the Votruba-Blokhintsev Programs.


\smallskip


\begin{thebibliography}{96}

\bibitem{C}
F.\,Calogero, J.\,Math.\,Phys. {\bf 10} (1969) 2191; {\bf 10} (1969) 2197.

\bibitem{OP}
M.A.~Olshanetsky, A.M.~Perelomov, Phys. Rept. {\bf 71}, 313 (1981);  {\bf 94}, 313 (1983).

\bibitem{C2}
A.P.\,Polychronakos, J.\,Phys.\,A:\,Math.\,Gen. {\bf 39} (2006) 12793.

\bibitem{AFF}
V.\,de\,Alfaro, S.\,Fubini, G.\,Furlan,
Nuovo Cim. {\bf A34} (1976) 569.

\bibitem{BH}
P.\,Claus, M.\,Derix, R.\,Kallosh, J.\,Kumar, P.K.\,Townsend, A.\,Van\,Proeyen, \\
Phys. Rev. Lett. {\bf 81} (1998) 4553.

\bibitem{GT}
G.W.\,Gibbons, P.K.\,Townsend, Phys. Lett. {\bf B454}
(1999) 187.

\bibitem{MS}
J.\,Michelson, A.\,Strominger,
Commun. Math. Phys. {\bf 213} (2000) 1; JHEP {\bf 9909} (1999) 005; \\
A.\,Maloney, M.\,Spradlin, A.\,Strominger, JHEP {\bf 0204} (2002) 003.

\bibitem{FM}
D.Z.\,Freedman, P.F.\,Mende, Nucl.Phys. {\bf B344}, 317 (1990).

\bibitem{Vas}
L.\,Brink, T.H.\,Hansson, M.A.\,Vasiliev, Phys. Lett. {\bf B286} (1992) 109; \\
L.\,Brink,
T.H.\,Hansson, S.\,Konstein, M.A.\,Vasiliev, Nucl. Phys. {\bf B401} (1993) 591.

\bibitem{W}
N.\,Wyllard, J.\,Math.\,Phys. {\bf 41} (2000) 2826.

\bibitem{BGK}
S.\,Bellucci, A.\,Galajinsky, S.\,Krivonos, Phys.\,Rev. {\bf D68} (2003) 064010.

\bibitem{BGL}
S.\,Bellucci, A.V.\,Galajinsky, E.\,Latini, Phys.\,Rev.
{\bf D71} (2005) 044023.

\bibitem{GLP}
A.\,Galajinsky, O.\,Lechtenfeld, K.\,Polovnikov, Phys. Lett. {\bf B643} (2006) 221;\\
JHEP {0711} (2007) 008; JHEP {\bf 0903} (2009) 113.

\bibitem{BKS}
S.\,Bellucci, S.\,Krivonos, A.\,Sutulin,
Nucl.Phys. {\bf B805} (2008) 24.

\bibitem{KLP}
S.\,Krivonos, O.\,Lechtenfeld, K.\,Polovnikov, Nucl. Phys. {\bf B817} (2009) 265.

\bibitem{DI}
F.\,Delduc, E.\,Ivanov, Nucl. Phys. {\bf B753}
(2006) 211, {\bf B770} (2007) 179.

\bibitem{ChS}
L.\,Faddeev, R.\,Jackiw,
Phys. Rev. Lett. {\bf 60} (1988) 1692;\\
G.V.\,Dunne, R.\,Jackiw, C.A.\,Trugenberger,
Phys. Rev. {\bf D41} (1990) 661;\\
F.\,Roberto, R.\,Percacci, E.\,Sezgin, Nucl. Phys. {\bf B322} (1989) 255;\\
P.S.\,Howe, P.K.\,Townsend,
Class. Quant. Grav. {\bf 7}
(1990) 1655.

\bibitem{Poly0}
A.P.~Polychronakos,
 Phys. Lett. {\bf B266}, 29 (1991).

\bibitem{FIL1}
S.\,Fedoruk, E.\,Ivanov, O.\,Lechtenfeld,
Phys. Rev. {\bf D79} (2009) 105015.

\bibitem{FIL2}
S.\,Fedoruk, E.\,Ivanov, O.\,Lechtenfeld, JHEP {\bf
0908} (2009) 081.

\bibitem{FIL3}
S.\,Fedoruk, E.\,Ivanov, O.\,Lechtenfeld, {\it New $D(2,1; \alpha)$ Mechanics with Spin
Variables}, \\ {\tt arXiv:0912.3508 [hep-th]}.

\bibitem{Gorsky}
A.~Gorsky, N.~Nekrasov,
Nucl. Phys. {\bf B414} (1994) 213.

\bibitem{IL}
E.\,Ivanov, O.\,Lechtenfeld,
JHEP {\bf 0309} (2003) 073.

\bibitem{GIOS}
A.S.\,Galperin, E.A.\,Ivanov, V.I.\,Ogievetsky, E.S.\,Sokatchev, {\it Harmonic Superspace}, \\
Cambridge Univ. Press, 2001.

\bibitem{Poly1}
A.P.~Polychronakos, JHEP {\bf
0104} (2001) 011; \\
B. Morariu, A.P.~Polychronakos, JHEP {\bf
0107} (2001) 006; Phys. Rev. {\bf D72} (2005) 125002.

\bibitem{IN}
E.~Ivanov, J.~Niederle,
Phys. Rev. {\bf D80} (2009) 065027.



\end{thebibliography}
\end{document}